\documentclass[12pt]{article}
\pdfoutput=1
\usepackage{graphicx}
\usepackage{color}
\usepackage{amssymb}
\usepackage{epstopdf}
\usepackage{subfigure}
\usepackage{appendix}
\usepackage{dtklogos}
\usepackage{tgpagella}
\usepackage{anyfontsize}
\usepackage{float}
\usepackage{amsmath,amsfonts, amssymb}
\usepackage{comment}
\newcommand{\be}{\begin{equation}}
\newcommand{\ee}{\end{equation}}
\newcommand{\ba}{\begin{eqnarray}}
\newcommand{\ea}{\end{eqnarray}}

\DeclareMathOperator{\Tr}{Tr}
\newcommand{\Cov}{\mathrm{Cov}}
\begin{document}
\begin{center} {\large{{\bf Monte Carlo Simulation  }}}
\end{center}
\begin{center} {\large{{\bf Of}}}
\end{center}
\begin{center} {\large{{\bf The Adjoint Coulomb Gas }}}
\end{center}
\vskip 0.5cm
\renewcommand{\thefootnote}{\fnsymbol{footnote}}
\centerline{
\textsc{ Omid Saremi}\footnote{omid@phas.ubc.ca}}
\vskip 0.5cm
\centerline{ Department of Physics and Astronomy, }
\centerline{University of British Columbia,}
\centerline{Vancouver, BC}

%%%%%%%%%%%%%%%
\begin{abstract}
\hrulefill

Monte Carlo simulation results for unitary matrix quantum mechanics, describing two-dimensional Yang-Mills theory coupled to a finite density of non-dynamical quarks (adjoint Coulomb gas), are presented. We characterize the deconfining transition in this model, by measuring the Polyakov Loop Susceptibility and employing finite-size scaling analysis. We provide evidence that the phase transition is first-order. Our results are consistent with the outcome of earlier large-$N$ studies of the model.

\end{abstract}

%%%%%%%%%%%%%%%%
\section{Introduction}
%%%%%%%%%%%%%%%%
Understanding the phase structure of QCD at finite temperature and density, in particular the so-called confinement-deconfinement transition, is an important subject with numerous applications, ranging from the interpretation of data generated at RHIC to understanding neutron stars \cite{Fukushima}.

Establishing the existence and characterizing properties of the QCD critical behavior are often beyond the realm of conventional perturbation theory. However, there are exceptions to this; in certain model gauge theories (pure or coupled to matter) the critical behavior can be brought under perturbative control, for instance in large-$N$ gauge theories defined on compact spaces \cite{Aharony1, Aharony2} or on $R^3\times S^{1}$ \cite{Mithat1, Mithat2}. 

Although, in recent years AdS/CFT has provided a quantitative window into the non-perturbative dynamics of certain model gauge theories \cite{Mateos}, the only practical and non-perturbative definition of QCD is still in terms of lattice field theory. Although there has been attempts to define field theories in terms of their transseries \cite{Mithat3}.
Lower dimensional toy models could serve as useful starting points. Recall that as the number of space-time dimensions increase, lattice simulations tend to be on smaller lattices, therefore do not allow for precise continuum extrapolations. A prohibitive computational cost also amounts to worse statistics. With the increase in Monte Carlo uncertainties, simulation results become less definitive. 

Yang-Mills theory in two-dimensions at finite temperature $T$, coupled to a density of non-dynamical adjoint color sources is an interesting playground for studying the phase structure of QCD-like theories. This system, coined as the ``adjoint Coulomb gas'' is the non-abelian cousin of the Coulomb gas \cite{Gordon1}. This model is interesting from another angle: It turns out \cite{Gordon2} that the adjoint Coulomb gas is effectively a \rm{unitary} matrix quantum mechanics (UMQM), defined on an infinite line (in the sense of quantum mechanics defined on the space of unitary matrices). ``Hermitian'' matrix quantum mechanics has already appeared in the context of the $D0$-brane black holes/matrix quantum mechanics duality \cite{Itzhaki}. In recent years, it has garnered even more attention by becoming one of the first systems in which a test of a strong-weak duality (like AdS/CFT) has been attempted from the strongly coupled side \cite{Hanada, Catterall1, Catterall2}. 
   
In the present work, we report on Monte Carlo simulation results performed on the UMQM description of the adjoint Coulomb gas. We mainly focus on its thermodynamics and critical behavior, at moderately large number of colors $N$. The analysis here is meant to be complementary to the large-$N$ studies performed in \cite{Gordon1, Gordon2}, but using a more versatile tool like Markov chain Monte Carlo.  

We utilize finite-size scaling analysis (FSS) to locate the deconfining transition in finite volume \cite{Engels1, Engels2}. We extrapolate our finite system size Monte Carlo results to infinite volume. Based on ``action histograms'' and the FSS, we provide evidence that the deconfining transition in this model is {\it first-order}. This is in line with \cite{Gordon1}.

There is another, perhaps more unorthodox, reason why this system is studied here using Monte Carlo: In this model, fugacity for the external quarks can be made a non-uniform function of the spatial extent of the system. It can then serve as a toy model, where effects of ``disorder'' in a gauge theory can be investigated. In such situations, traditional tools like perturbation theory or the large-$N$ are of little use.

The present paper is organized as follows. In section two we describe the setup and highlight some relevant known results. In section three lattice action and measured observables are discussed. Section four contains our simulation results on the deconfining transition and critical behavior of the Polyakov Loop Susceptibility. In section five the nature of the transition is determined using a variety of methods. We finally conclude with a summary. To make our simulations and the corresponding statistical data analysis reproducible, detailed accounts of our parameter choices are included as well. Some of the techniques used in this paper are briefly reviewed in an appendix.
%%%%%%%%%%%%%%%%% 	
\section{The model}
%%%%%%%%%%%%%%%%%

Consider the partition function of a two-dimensional $U(N)$ Yang-Mills theory coupled to a fixed number  $M$ of non-dynamical quarks, sitting at positions $X=\{x_{i}|~i=1\cdots M\}$ and in contact with a heat bath at temperature $T=1/\beta$\footnote{Later on, we consider the grand canonical ensemble. The mass dimensions for the gauge coupling and field are $[g_{YM}]=[A]=1$}
\be\label{model}
Z=\int \prod_{\mu\in\{\tau, x\}}[d A_{\mu}] e^{-\frac{1}{2g^2_{YM}}\int_{0}^{\beta}d\tau\int dx~ \Tr F^2 }\prod_{i=1}^{M}\Tr\mathcal{P}\exp(i\int_{0}^{\beta} d\tau A^{adj}_{\tau}(\tau, x_i)),
\ee
where by superscript $adj$, we indicate that all external quarks are taken to be in the adjoint representation of the gauge group. This system is rightfully called the ``adjoint Coulomb gas'' in $D=1$ (spatial) dimension and finite temperature $T$ \cite{Gordon1}. In the continuum, it describes (1+1)-dimensional QCD coupled to heavy adjoint matter.
 
As alluded to earlier, this model is equivalent to a UMQM. The complete and more rigorous derivation of the UMQM from (\ref{model}) is given in \cite{Gordon1, Gordon2} and will not be reproduced here. Instead, the steps are briefly sketched. In the Hamiltonian picture, the physical spectrum of (\ref{model}) consists of those energy eigenstates of the Hamiltonian $\hat{H}$, which are further annihilated by the Gauss' s constraint. As usual, the partition function at finite temperature $T$ and fixed number of quarks is given by $Z=\Tr(e^{-\beta\hat{H}})$, where the trace is taken over all physical states. Projection onto gauge invariant states can be performed formally by employing Gross, Pisarski and Yaffe representation \cite{Gross}
\be\label{model2}
Z(\beta, \{x_1\cdots x_M\})= \int [dA][dU]\langle A|e^{-\beta \hat{H}}|A^{U}\rangle \prod_{i=1}^{M}\Tr U^{adj}(x_i),
\ee
where $A^{U}=UAU^{\dagger}-iU\nabla U^{\dagger}$ and $U(x_i)$ are gauge group elements in the adjoint representation.\footnote{The idea is to gauge transform the state at one side of the trace and then integrate over the space of gauge transformations to get a gauge invariant partition function.} The basis states $|A\rangle$ are the eigenstates of the operator $\hat{A}^{a}(x)$ ($a$ denotes a gauge index), tensored with a convenient basis for the non-dynamical quarks. In transitioning to the grand canonical ensemble, where the number of quarks is only thermodynamically determined, one needs to integrate over the quark positions and multiply by powers of fugacity. Note that the non-dynamical quarks are distinguishable and classical. Formally, the grand canonical ensemble is given by\footnote{The fugacity $\lambda$ acquires a mass dimension equal to one from integrating over quark positions.} 
\be
Z_{G}(\beta)=\sum^{\infty}_{M=0}\int \prod^{M}_{i=1}dx_i~ \frac{\lambda^M}{M!}Z(\beta, \{x_1\cdots x_M\}).
\ee
This gives a gauged principal chiral model with a potential term \cite{Gordon1, Gordon2}
\be\label{umqm}
S_{eff}=\int^{\infty}_{-\infty} dx ~[\frac{N}{2\gamma}\Tr (DU(x) DU^{\dagger}(x))-\lambda \Tr U(x)\Tr U^{\dagger}(x)], \quad Z_{G}=\int [dA][dU] e^{-S_{eff}},	
\ee  
where 
\be
DU(x)=\partial_{x}U(x)+i[A_{x}(x),U(x)], \quad  \gamma=\frac{g^2_{YM} N}{2T},
\ee
and the known relation between trace in the fundamental and adjoint representations of $U(N)$ was used. The factor of $N$ in $\gamma$ is such that at large-$N$ both kinetic and potential terms are $\mathcal{O}(N^2)$, if $\gamma$ is kept fixed (in the same units as $\lambda$ by tuning $g_{YM}$). The saddle point approximation is therefore justified. Also note that in this model only $\lambda/\gamma$ is physically meaningful. The model (\ref{umqm}) is defined on an infinite line. With this spatial topology, there is no obstruction in gauge fixing to $A=0$ and decoupling $A$ entirely. In the following we work in this gauge. If one re-interprets space as Euclidean time 
\be
x\rightarrow -i\tau, \quad A_{x}\rightarrow iA_{x},
\ee
then model (\ref{umqm}) describes the Euclidean action of a quantum mechanical system defined on the space of unitary matrices. The fact that the $\tau$-direction will have infinite extent implies that the free energy density of the original model (\ref{model}) is the ground state energy of the UMQM in (\ref{umqm}). 
%%%%%%
\subsection{\small Large-$N$ Thermodynamics}
%%%%%%
In the large-$N$ limit, thermodynamics of (\ref{umqm}) can be treated analytically, using the collective field method \cite{Zarembo, Gordon1, Gordon2}. It turns out that in the $\lambda-\gamma $ plane, there is a phase boundary across which the free energy density jumps from $\mathcal{O}(1)$ to $\mathcal{O}(N^2)$.

There is a line in the $\lambda-\gamma$ plane, near which the uniform eigenvalue distribution (a hallmark of the confining phase) becomes unstable and higher harmonics start to develop. The first harmonic goes unstable below a line $\gamma_{unstable}(\lambda)=4\lambda$, which is found by studying the perturbative stability of the uniform eigenvalue density, using collective field theory. On the other hand, the jump in free energy density occurs on a slightly different line $\gamma=\gamma_{transit}(\lambda)=4.219$. The line on which the deconfining  phase terminates is $\gamma=\gamma_{*}(\lambda)=4.433\lambda$. 
Intuitively, the origin of this phase transition is easy to understand. The second term in the action (\ref{umqm}) is responsible for an attractive ``inter-eigenvalue potential''. At the same time, the Haar measure induces an entropic repulsion between the eigenvalues. The competition between the two leads to a strong first-order phase transition.\footnote{Note that the sign of the second term is important here. For a phase transition to exist an attractive force between eigenvalues is needed.} 

%We will compare the above $N=\infty$ results to our simulations. 
%Monte Carlo simulations are only done at finite volume and for finite $N$. Deducing and characterizing the phase transition requires FSS which we do. 
%%%%%%%%%%%%%%%%%%%
\section{Lattice action/Observables}
 %%%%%%%%%%%%%%%%%%%
We use the following continuum action

\be\label{continuum}
S=\sqrt{\frac{N\lambda}{2\gamma}}\int^{\infty}_{-\infty}dx~[\Tr (D_xUD_xU^{\dagger})-\Tr U\Tr U^{\dagger}],
\ee
where in this version of the action, the coordinate $x$ is dimensionless. Consider the following one-dimensional lattice $x_j=ja$ for $j=0\cdots K+1$. The relevant part of the lattice action is \footnote{There are constant pieces which we subtract. They drop out of the expected values.}
\be\label{lattice}
\sqrt{\frac{2\gamma}{N\lambda}}S_{lattice}=-\sum^{K}_{i=0}[\frac{1}{a}\Tr(U_{i}^{\dagger}U_{i+1}+U_{i+1}^{\dagger}U_{i})+a\Tr U^{\dagger}_{i}\Tr U_{i}],
\ee
where $a=L/(K+1)$ is the UV cutoff and $L$ is the dimensionless system size. A Dirichlet boundary condition is imposed at the endpoints by demanding $U_{0}=U_{K+1}=\mathbb{I}_{N\times N}$. For any fixed $L$, the continuum limit is simply achieved by $K\rightarrow \infty$. We also need to take $L\rightarrow \infty$. We employ Markov chain Monte Carlo method to simulate (\ref{lattice}). More precisely, the Metropolis-Hasting local update algorithm is used to estimate the observables of interest. This algorithm generates the Haar measure over $U(N)$ matrices automatically. 
 
In this work, our primary objective is to study the phase diagram of the continuum model (\ref{continuum}) as a function of $\lambda/\gamma$, for moderately large $N$ and $L$. From earlier large-$N$ studies, a first-order confinement-deconfinement transition is to be expected.
 
To detect phase boundaries, an order parameter is needed. The traditional order parameter for a deconfining transition is the Polyakov Loop $\langle \Tr U\rangle/N$: While it is zero  when the system confines, a non-vanishing expected value develops in the deconfinig phase. Here we use a closely related observable, i.e., the  expectation value of the quark density at finite temperature $T$ in units of $N^2$
 \be
 n=\frac{\lambda}{N^2L}\frac{\partial \ln{Z_{G}}}{\partial\lambda}=\frac{\lambda}{N^2L}\langle\int^{L}_{0} dx~ \Tr U(x) \Tr U^{\dagger}(x)\rangle, 
\ee
where $\lambda$ is the fugacity and $L$ is the system size. In the confining phase, $n$ is small while in the deconfined phase it jumps to $\mathcal{O}(1)$. See \cite{Gordon1}. The discretized version of the order parameter is 
\be
n=\frac{\lambda}{N^2K}\langle\sum^{K}_{i=1}\Tr (U_i)\Tr (U^{\dagger}_i)\rangle. 
\ee
The quark density $n$ has mass dimension equal to one. We set units of mass by fixing (the other dimensionful coupling)  $\gamma$ throughout our simulations to 4.6. To force the system to undergo a phase transition, $\lambda$ is varied only. 

Technically a phase transition is infinitely sharp only in the infinite-volume limit or when the number of degrees of freedom is infinite, for instance at large-$N$. To measure and characterize phase transitions in a finite volume (which is all one can do with Monte Carlo), besides the order parameter, studying the critical behavior of the fluctuations of the order is also important. In practice, one measures a ``good tracer'' of the transition at finite volume $L$. The infinite volume critical coupling $\lambda_c=\lambda_c(\infty)$ is then calculated by extrapolating to $L=\infty$. It turns out the critical coupling $\lambda=\lambda_c(L)$, where the Polyakov Loop Susceptibility (denoted by $\chi$) exhibits a peak, is a good tracer of the actual de-confinement transition at infinite volume \cite{Privman, Barber, LatticeBook}
\be
\frac{\chi}{K}=\langle |P^2|\rangle-\langle|P|\rangle^2,
\ee
where
\be\label{avgpolyloop}
P=\frac{1}{K} \sum^{K}_{i=0}\frac{1}{N} \Tr (U_i).
\ee
We present our measurements of $\chi$ and its extrapolation to infinite volume to compute $\lambda_c=\lambda_c(L=\infty)$ in section (\ref{chimeas}). 
%%%%%%%%%%%%%%%%%%%%
\section{Monte Carlo Simulation results} 
%%%%%%%%%%%%%%%%%%%%
%%%%%
\subsection{Behavior of the order parameter}
%%%%%
In this section, our Monte Carlo measurements of the quark density are presented. In order for the CPU time to remain reasonable, we limit ourselves to $N= 10, 12$ and system sizes $L=8, 13$. We also show our results for $N=14, L=25$ (with lower statistics). Larger simulations are completely within reach, although the performed simulations already capture the physics we are after in this note. 

\begin{figure}
\centering
\mbox{\subfigure{\includegraphics[width=3.0in]{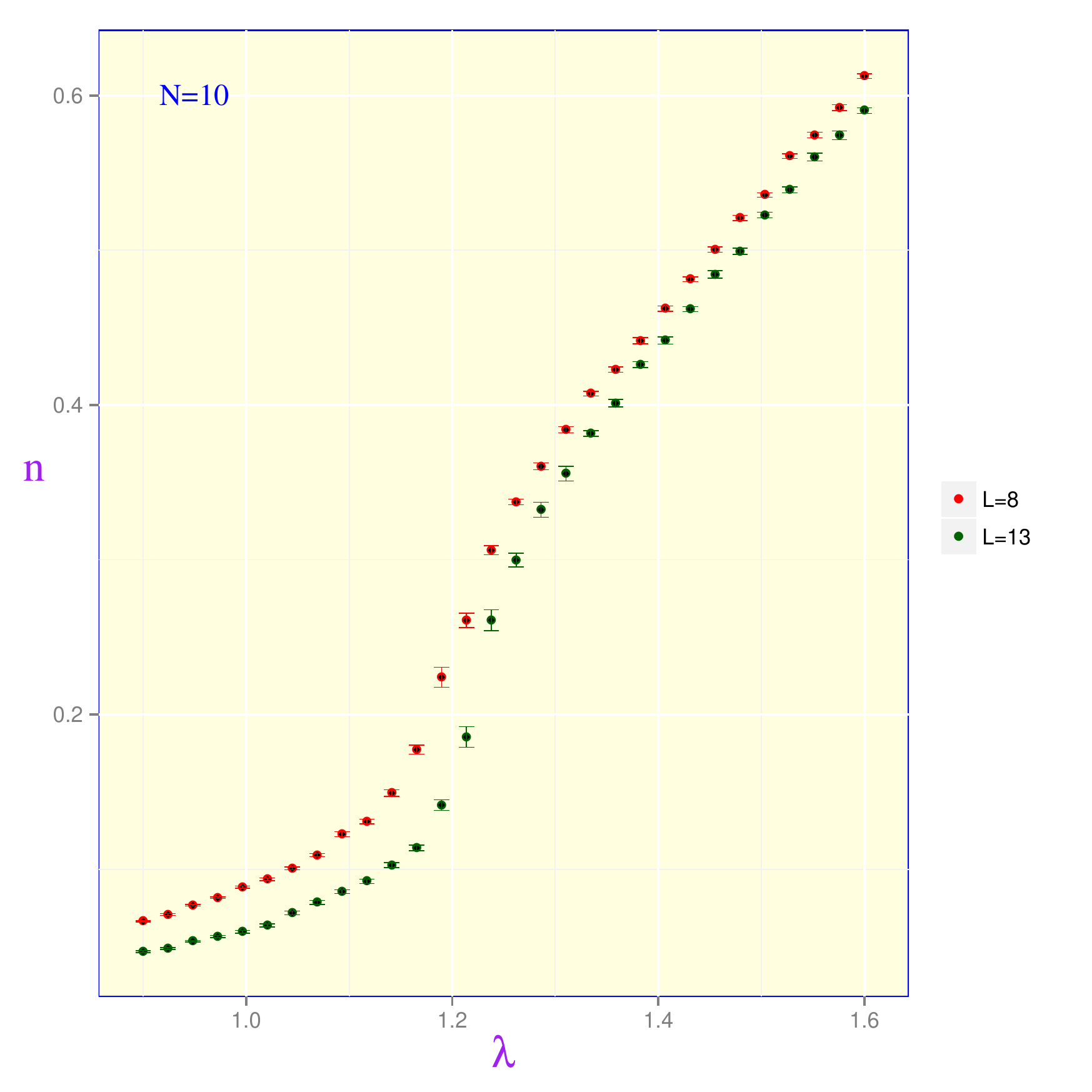}
\quad
\subfigure{\includegraphics[width=3.0in]{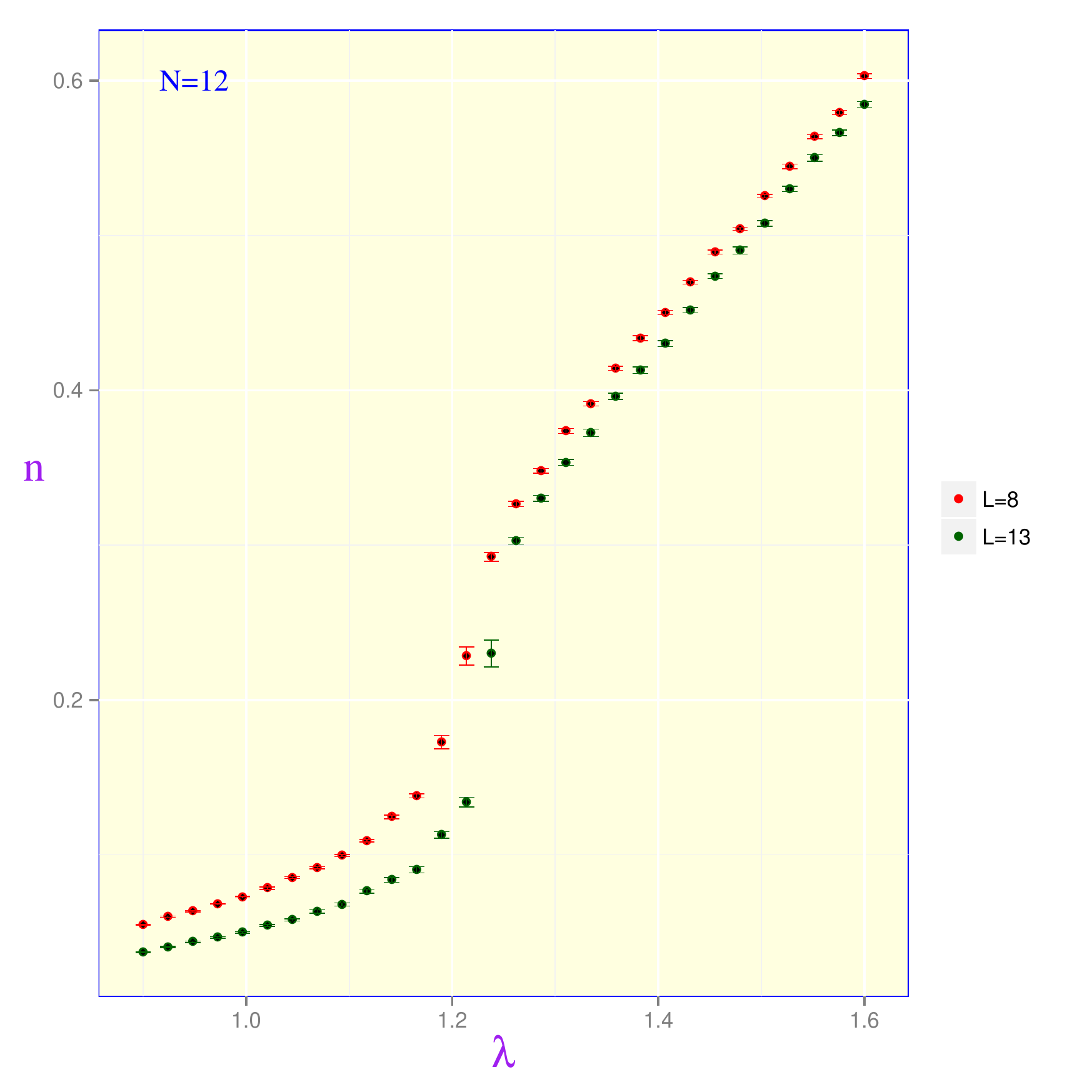} }}}
\caption{}
\label{fig:1}
\end{figure}

To generate plots in FIG.(\ref{fig:1}), $2\times 10^5$ measurements were performed. Between consecutive measurements, 300 intermediate Monte Carlo sweeps were discarded to reduce autocorrelations. To reach equilibrium, the first $10^5$ sweeps were thrown away (to reduce systematic errors from the transient out-of-equilibrium state). At the beginning of each run, a collection of $5\times10^4$ unitary matrices in the neighborhood of identity were constructed. These matrices were used to propose local updates at each site during sweeps (see the appendix). To respect the detailed-balance condition, half of these matrices were taken to be the Hermitian-conjugate of the remaining half. The Mersenne Twistor algorithm was used to generate pseudo random numbers. We tried a few different values of the lattice spacing. We show the results corresponding to the lattice spacing $a=0.1$ and coarser. Unitarity of the field configurations along the Markov chain were regularly checked. This was done by measuring a defined distance 
\be
\mathfrak{D}_i=\frac{1}{N^2}\sum_{k, \ell} |(U_iU_i^{\dagger})_{k\ell}-\mathbb{I}_{k\ell}|,
\ee
between $U_iU_i^{\dagger}$ and the identity matrix, for every lattice site $i$, every several Monte Carlo sweeps. The field configuration at a given site with a distance below a fixed threshold (set to be $10^{-10}$ in our simulations), was declared unitary. This is necessary, since at late Monte Carlo times accumulated rounding errors can destroy the unitarity property of the field configurations. If needed, a ``nearly'' unitary configuration was re-unitarized using the QR-algorithm. 

\begin{figure}[H]
\centering
\includegraphics[width=3.5in]{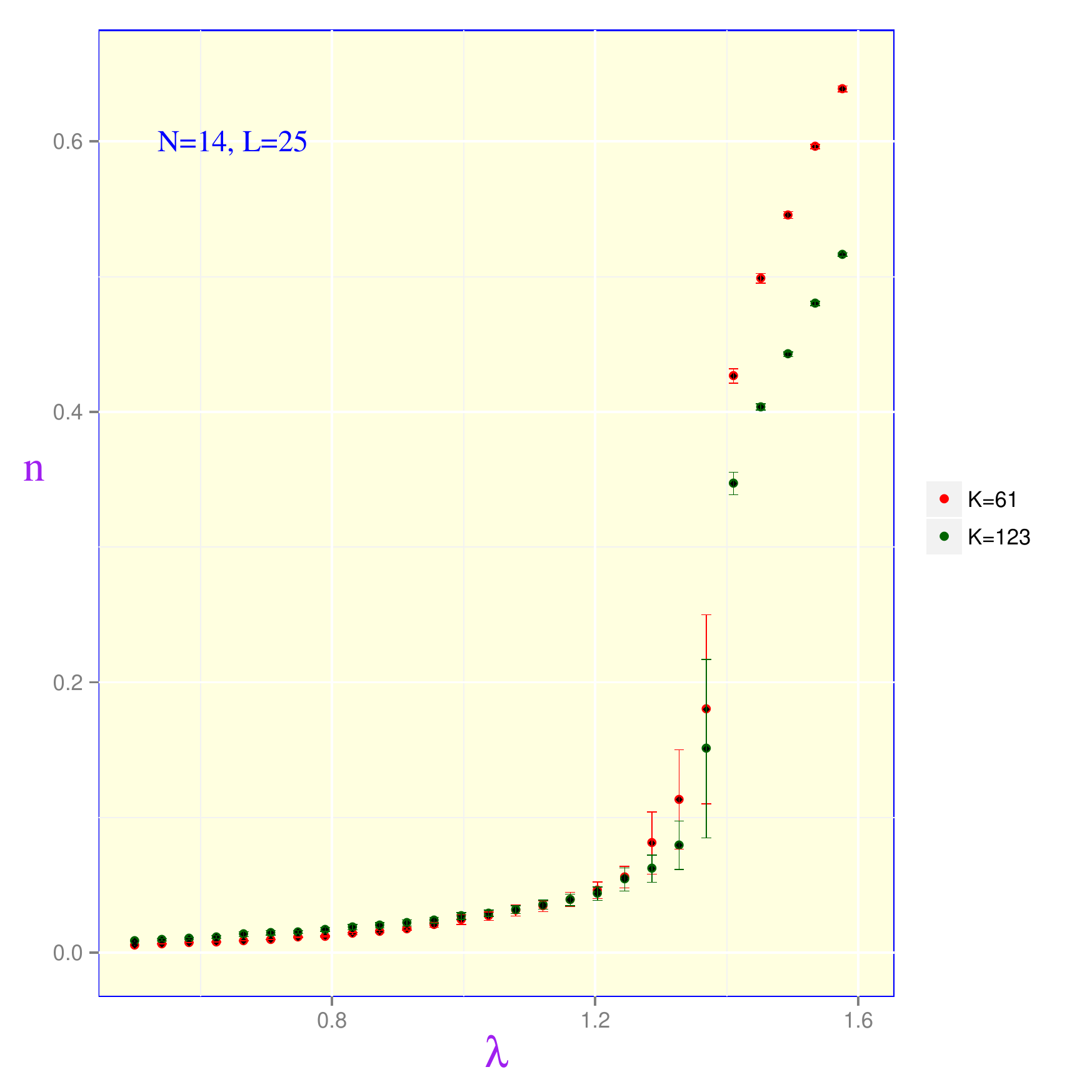}
\caption{}
\label{fig:2}
\end{figure}

As clearly seen from FIG.(\ref{fig:1}), there exists a rapid increase in the quark density at $\lambda=\lambda_c\approx 1.2$. Note that for a fixed $N$ (or $L$), the transition looks sharper as $L$ (or $N$) is increased. As emphasized before, $\lambda_c$ is subject to finite $N$, $L$ and $a$ corrections. In FIG.(\ref{fig:2}), the order parameter is plotted for $N=14$ and $L=25$. The number of Monte Carlo measurements taken was lower: For the dark red data points $4.5\times 10^4$ measurements (with 50 sweeps discarded in between) were taken, while for the red data points, $1.5\times10^4$ measurements with 30 sweeps discarded was used in the data analysis and error-bar estimates. The rapid surge in the quark density is even more clear here. Error-bars were computed by monitoring the autocorrelation times and binning the Monte Carlo time series. See the appendix. Lower statistics and longer autocorrelation times are the reasons for larger error-bars in the critical region. Note that the red data points belong to a run with a coarse lattice (with $K=62$), so we are further away from the continuum compared to the case $K=121$. The critical $\lambda$ seems not to care about this difference in lattice sizes and to be around $\lambda=\lambda_c\approx1.37$. In the following sections, $\lambda_c(L=\infty)$ will be determined.
 
%%%%%%%
\subsection{\small Polyakov Loop Susceptibility (PLS)/Zooming into the critical region}\label{chimeas}
\begin{figure}[H]
\centering
\mbox{\subfigure{\includegraphics[width=2.5in]{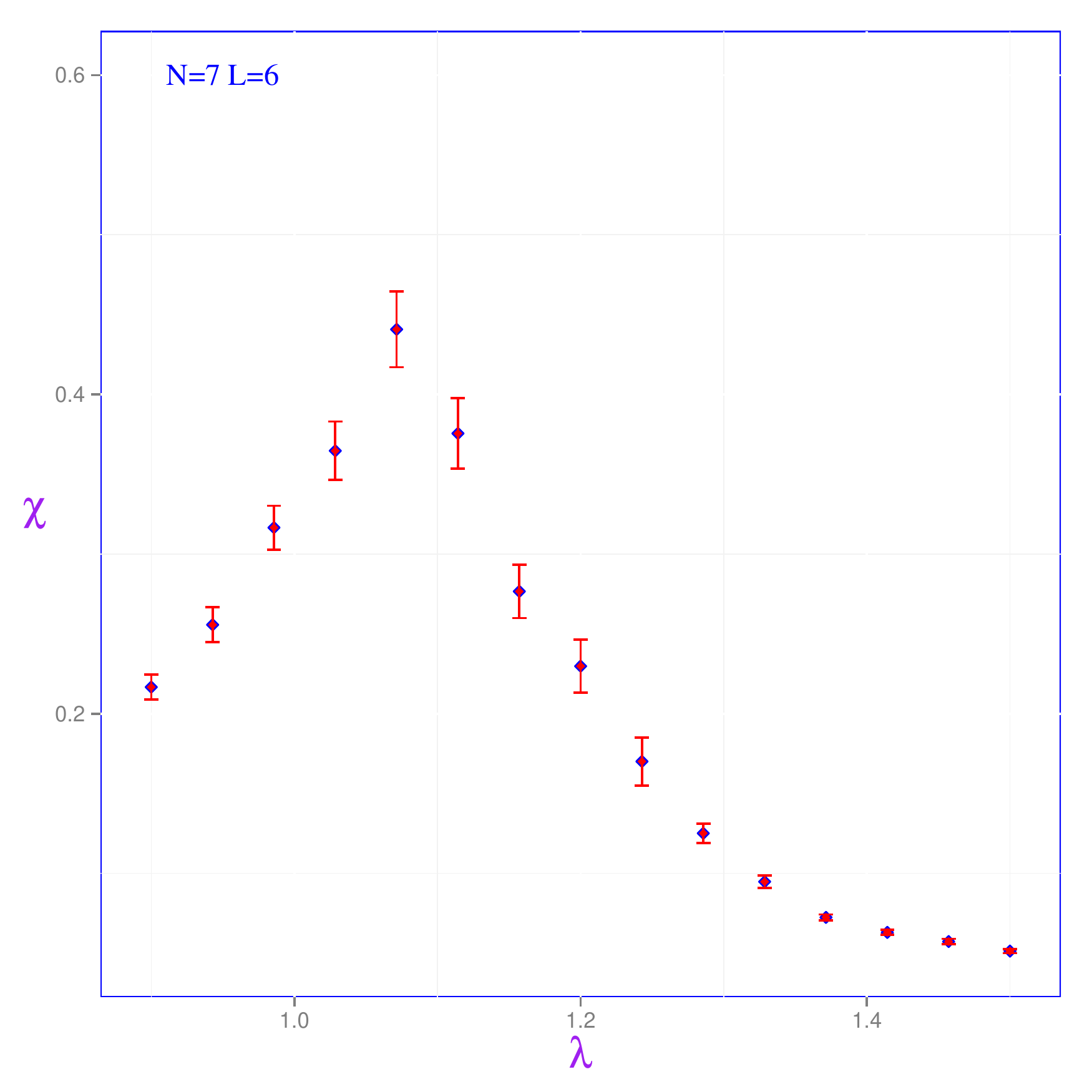}
\quad
\subfigure{\includegraphics[width=2.5in]{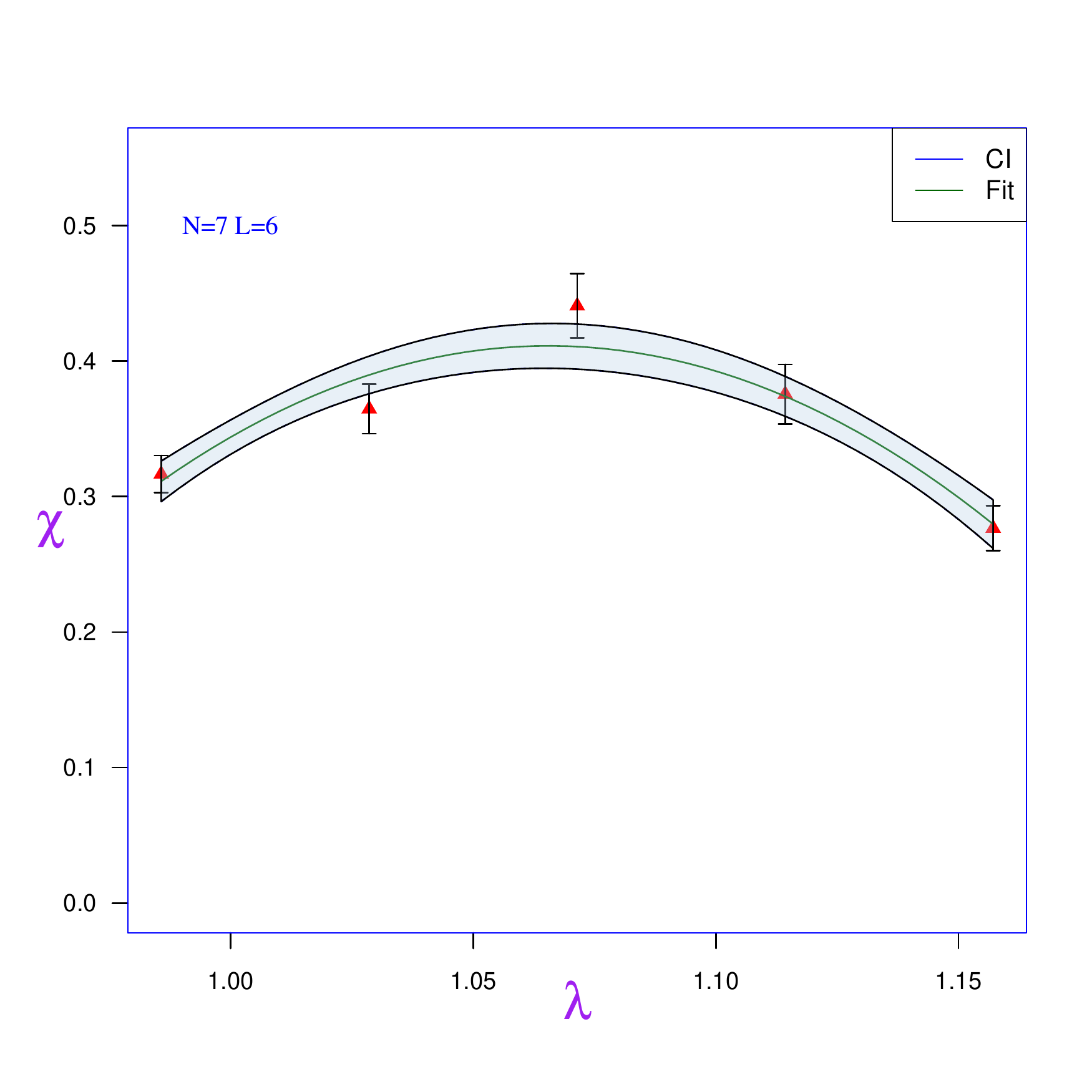} }}}
\caption{}
\label{fig:3}
\end{figure}

%%%%%%%
In this section we present our simulation results on susceptibility near the critical point. In FIG.(\ref{fig:3}), the PLS is plotted versus $\lambda$ for $N=7$ and $L=6$.  Next to it, is  the plot of the magnified region in the vicinity of the peak (mind the range on the $\lambda$-axis). To estimate the location of the maximum of $\chi$, we need to know $\chi$ as a continuous function of $\lambda$ in the critical region. This is usually done using ``reweighting'' the Monte Carlo data \cite{Fer}.\footnote{Re-weighting technique is usually used to extrapolate the Monte Carlo data taken for one value of a parameter to compute the estimates for the observable of interest for nearby values of that parameter, using action histograms.} We took an alternative route here. Near the peak we model the Monte Carlo data by a quadratic form $\chi=\alpha\lambda^2+\beta\lambda+\eta$. The maximum likelihood estimate (Chi-Square fit) of the regression coefficients  $\alpha$, $\beta$ and $\eta$ was then found. The colored band around the fitting curve (dark green) in FIG.(\ref{fig:3}) is the confidence band for the regression curve. It was determined by finding the prediction error of the regression curve. See the appendix for details.

\begin{figure}[H]
\centering
\mbox{\subfigure{\includegraphics[width=2.5in]{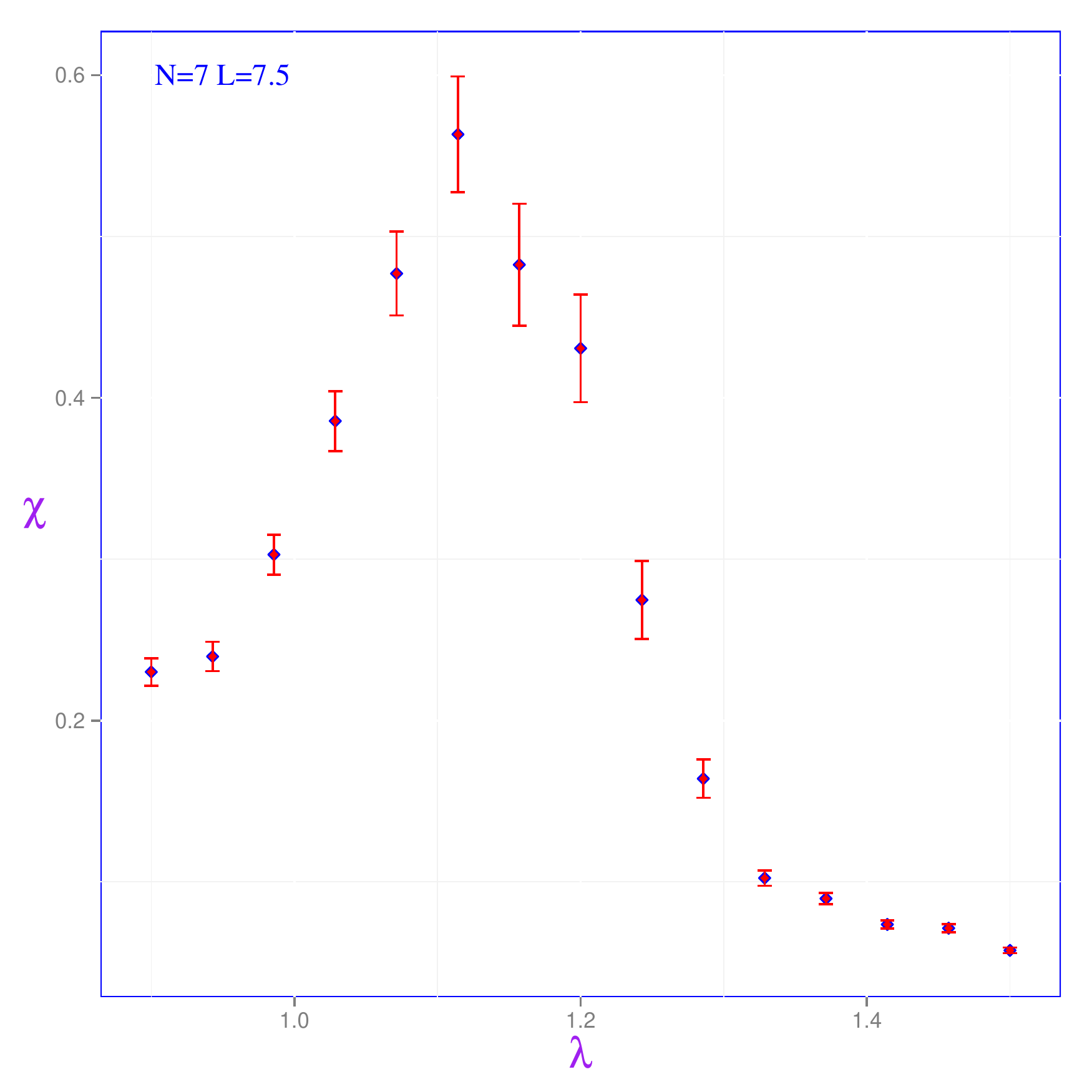}
\quad
\subfigure{\includegraphics[width=2.5in]{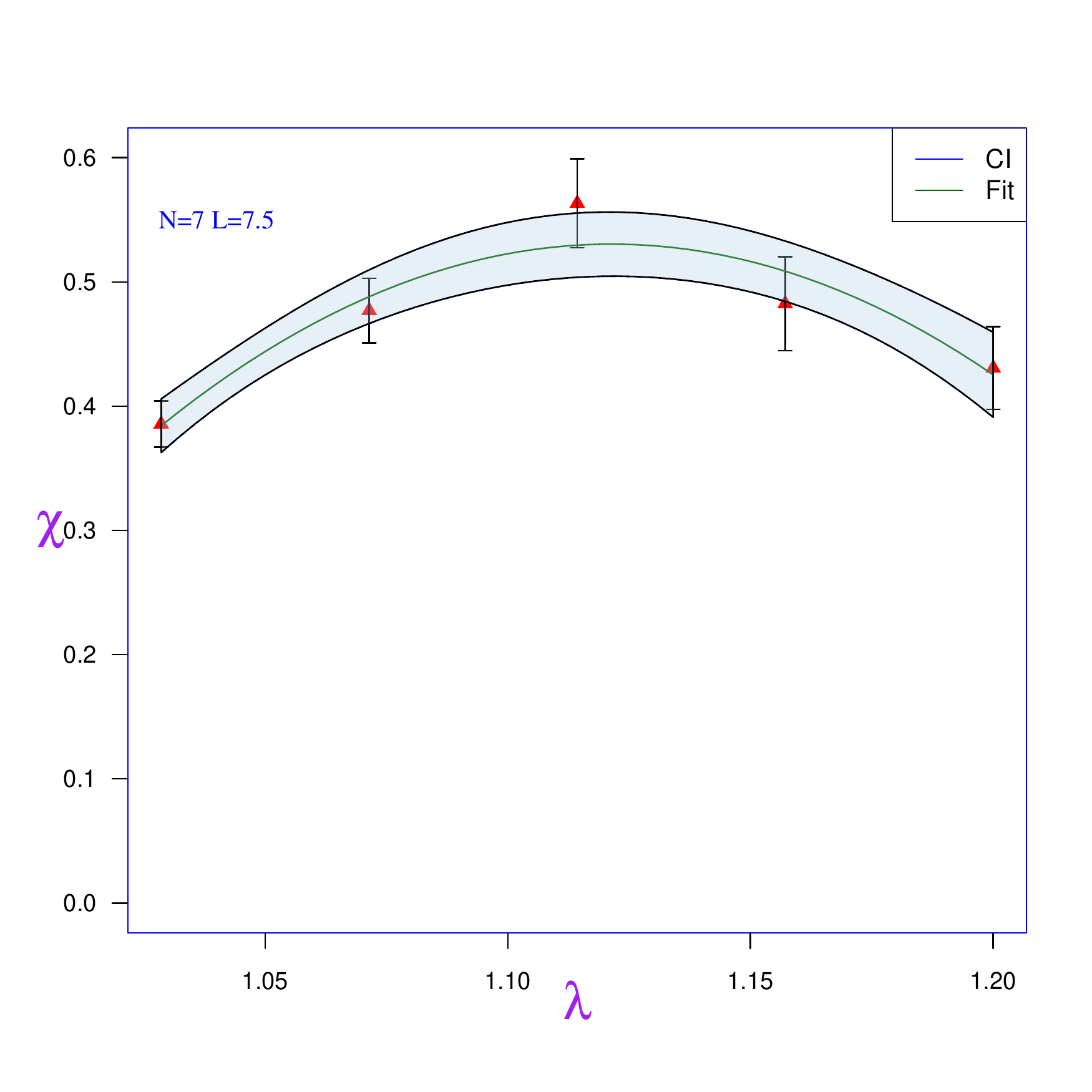} }}}
\caption{}
\label{fig:4}
\end{figure}

\begin{figure}[H]
\centering
\mbox{\subfigure{\includegraphics[width=2.5in]{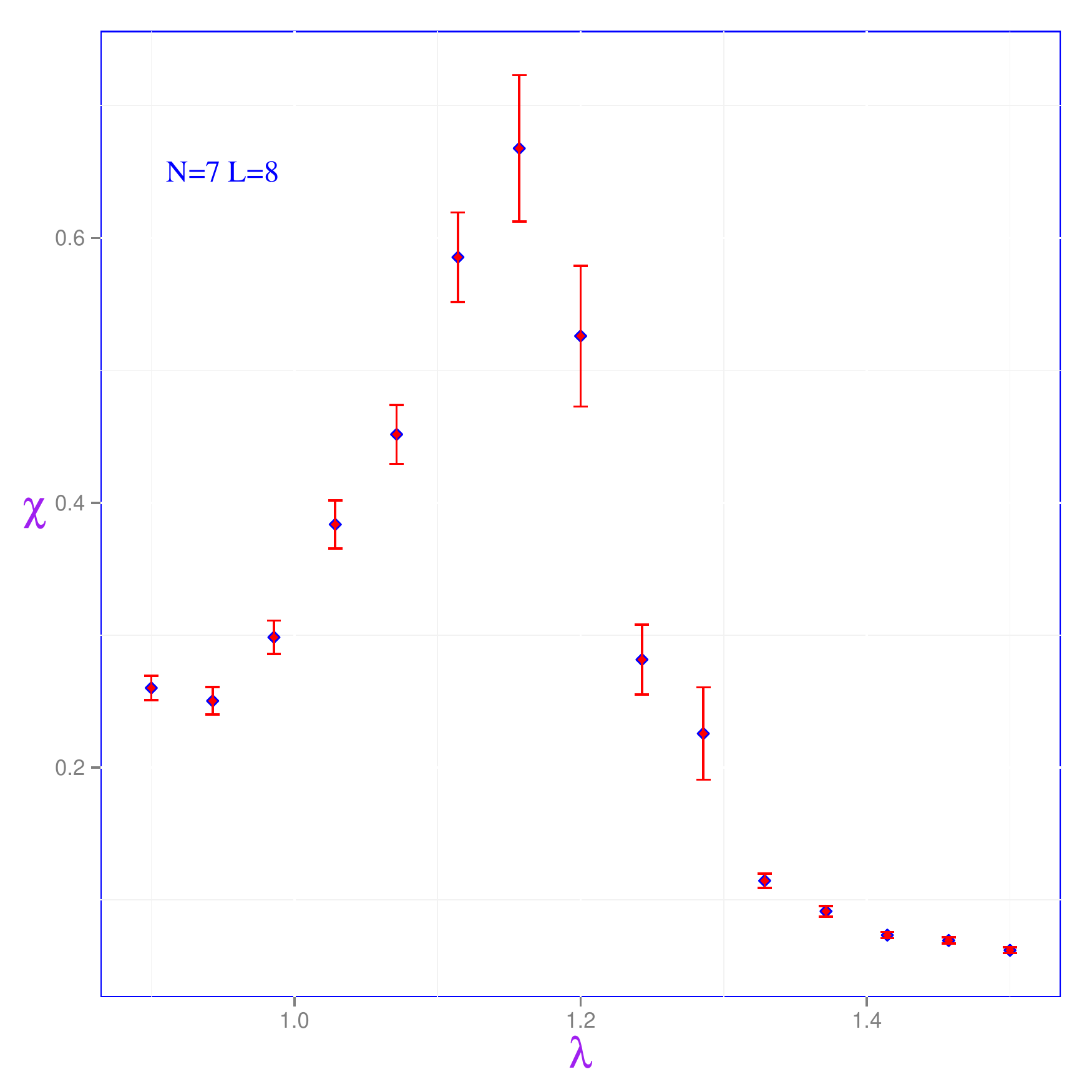}
\quad
\subfigure{\includegraphics[width=2.5in]{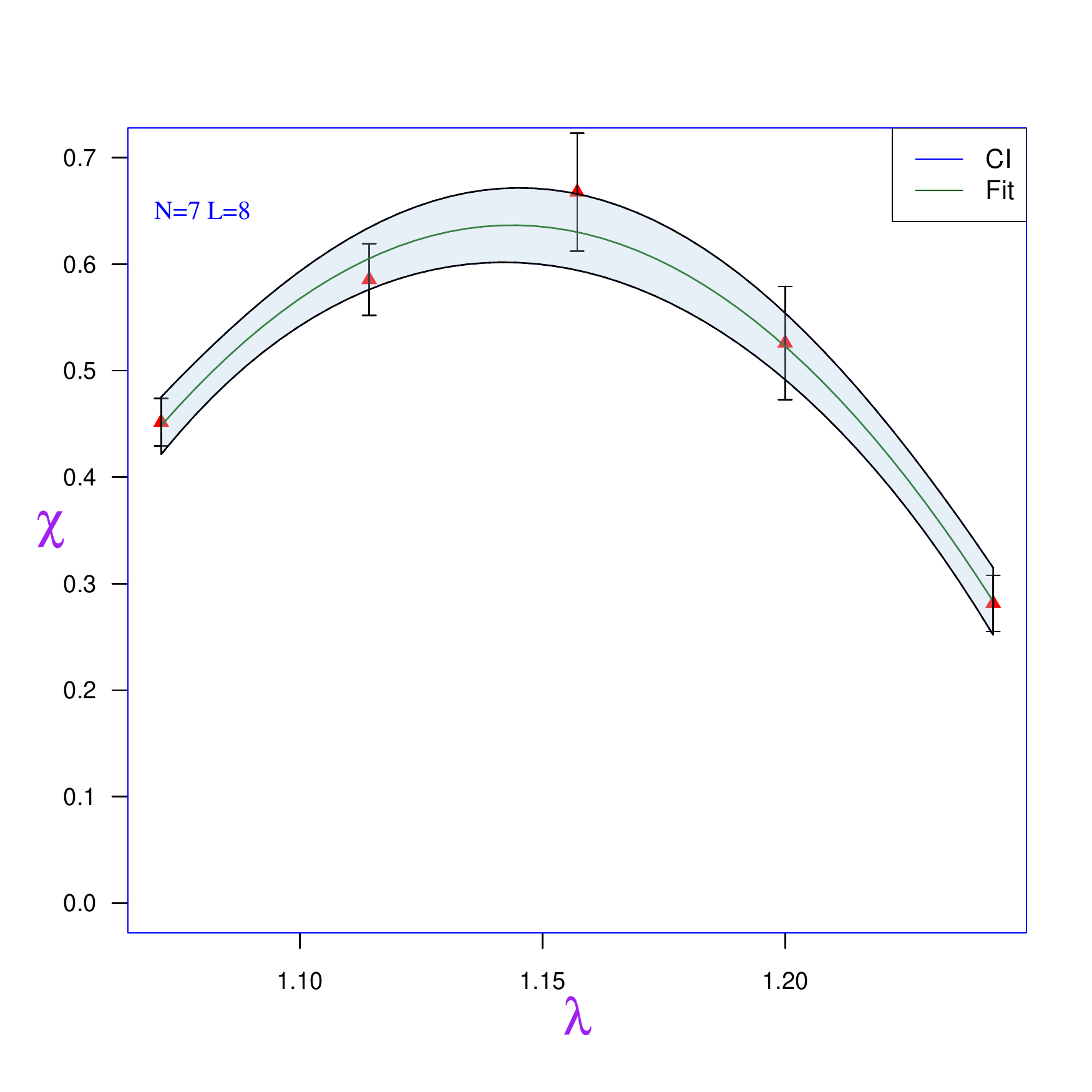} }}}
\caption{}
\label{fig:5}
\end{figure}

In FIG.(\ref{fig:4}) and FIG.(\ref{fig:5}) the PLS measurements for larger systems are displayed. We obtained three different values for the critical $\lambda$ for each $L$, by recording the maximum of the fitting curve and the two boundaries of the confidence region. In the next section this data is used to compute $\lambda_c$ in the infinite volume and to estimate its uncertainty. 

We did run simulations with higher values of $N$ and $L$, however autocorrelations in Monte Carlo time series start to diverge in the critical regime as the system size and/or $N$ grow. This leads to large uncertainties in the Monte Carlo estimates. For higher values of $N$ and $L$, the order parameter data was considerably less noisy. That is why we have gone up to $N=14$ and $L=25$. In order to have sufficiently tight estimates of $\chi$ at higher values of $N$ and $L$, longer computer time is necessary. In addition, at the level of algorithms, improvements need to be made to counter the critical increase in the autocorrelations.  

\begin{figure}[H]
\centering
\includegraphics[width=3.0in]{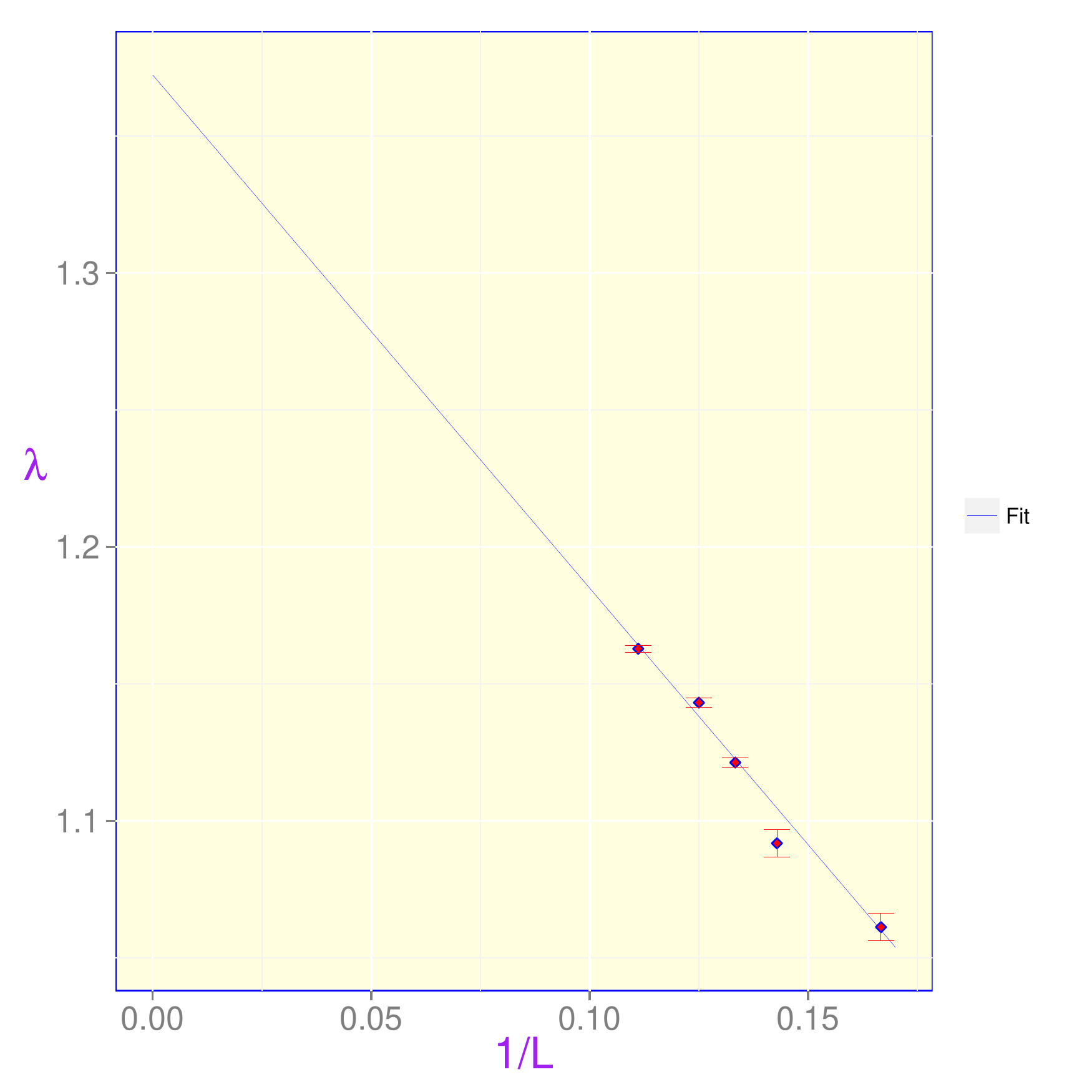}
\caption{}
\label{fig:6}
\end{figure}

Based on finite-size scaling theory, $\lambda_c$ is expected to exhibit the following behavior in the large-$L$ limit, if the transition is first-order \cite{Barber, Privman}
\be
\lambda_c=\lambda_c(\infty)+\frac{b_2}{L}+\cdots,
\ee
where dots represent higher order terms in $1/L$. This expected behavior is consistent with our simulation data, as evident in FIG.(\ref{fig:6}). Overlaid is the regression line fitted to the data (with $\chi^2=1.23$). The intercept $ \lambda_c=1.372\pm0.008$ is the predicted critical coupling in the infinite volume. The slope is $b_2=-1.87\pm 0.07$. Again, note that both the intercept and slope are subject to finite-$N$ and $a$ corrections: An inspection of the quark density data points for $N=14$ and $L=25$ in FIG.(\ref{fig:2}) reveals that $\lambda_c\approx 1.37$. This is reassuringly consistent with our finite-size analysis done at noticeably lower values of $N$ and $L$. This also indicates that any remaining discrepancy with the $N, L=\infty$ case, should mostly come from deviations from the continuum limit. 
Our preliminary results show that these corrections are consistent with 
\be
\lambda_c(N, L, a)=\lambda_c+\frac{b_1}{N^2}+\frac{b_2}{L}+b_3 a+\mathcal{R}(\frac{1}{N^2}, \frac{1}{L}, a),
\ee
where $\mathcal{R}$ represent the higher-order corrections. In this work we have computed $b_2$ and $\lambda_c$. A more extensive analysis is underway which will accurately determine the remaining coefficients $b_1$ and $b_3$.  
%%%%%%%%%%%%%%%%%%%%
%%%%%%%%%%%%%%%%%%%%
\section{Order of the transition}
%%%%%%%%%%%%%%%%%%%%
There are various ways of determining the order of a phase transition at finite volume. One common method is to look at the Monte Carlo time series itself. In fact, in the vicinity of a first-order transition, there will be tunneling events back and forth between the two competing phases. This behavior manifests itself as a time series, in which the observable spends most of its time taking one of the two values, with sharp transitions in between as Monte Carlo time progresses. These tunneling events should become increasingly rare as system size (or $N$) increases. The second (related) method is to inspect the action histograms near the transition. A double-peak structure is the definitive signature of a first-order transition.\footnote{This approach is not useful if the difference in the actions of the two phases is too small compared to the typical spread. Looking at histograms of the modulus of $P$, defined in (\ref{avgpolyloop}) is another alternative.} Finally, the third most common approach is to observe a latent heat which survives the infinite volume limit. These common methods have been employed\ before. See for instance \cite{Lucini}.

\begin{figure}[H]
\centering
\mbox{\subfigure{\includegraphics[width=3.0in]{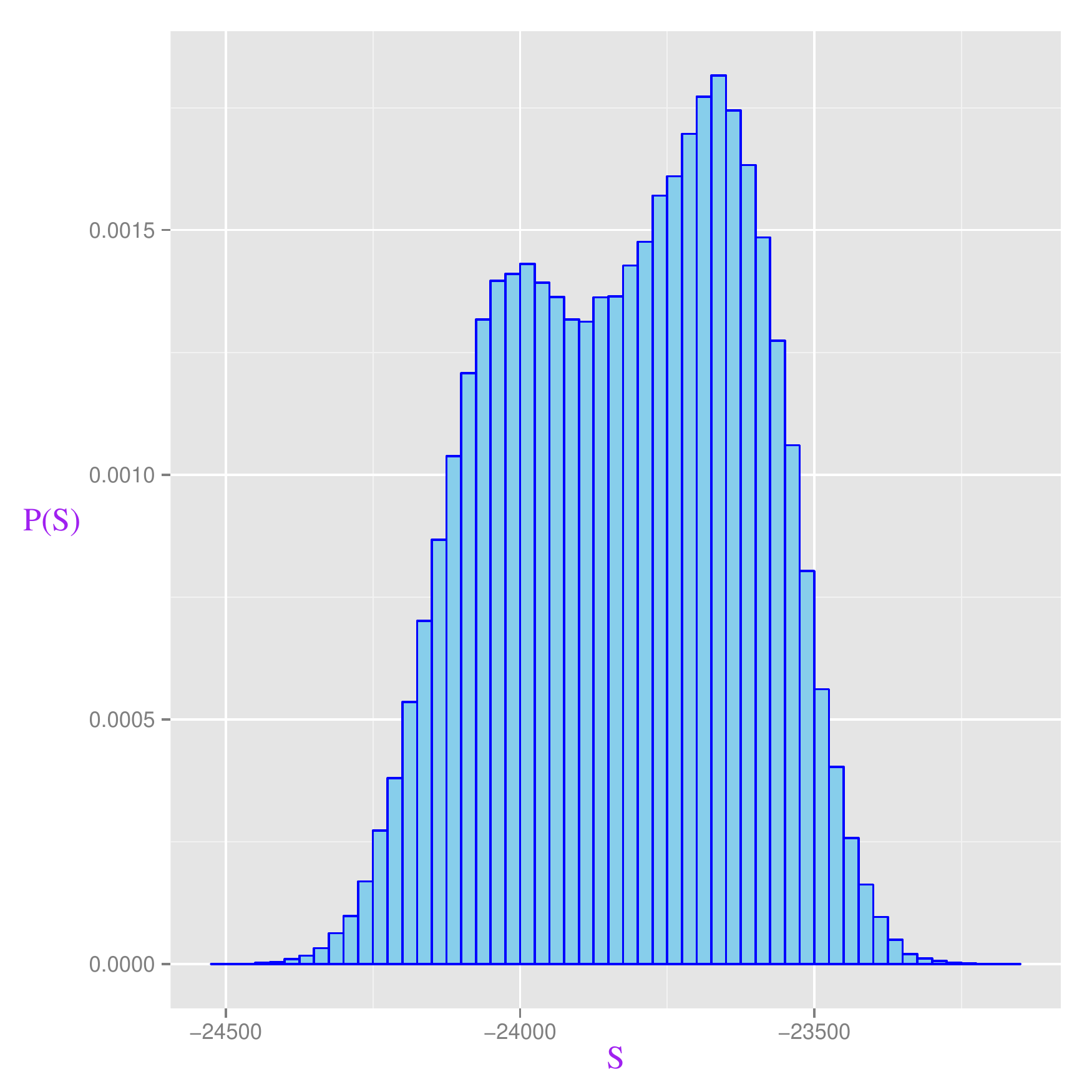}
\quad
\subfigure{\includegraphics[width=3.0in]{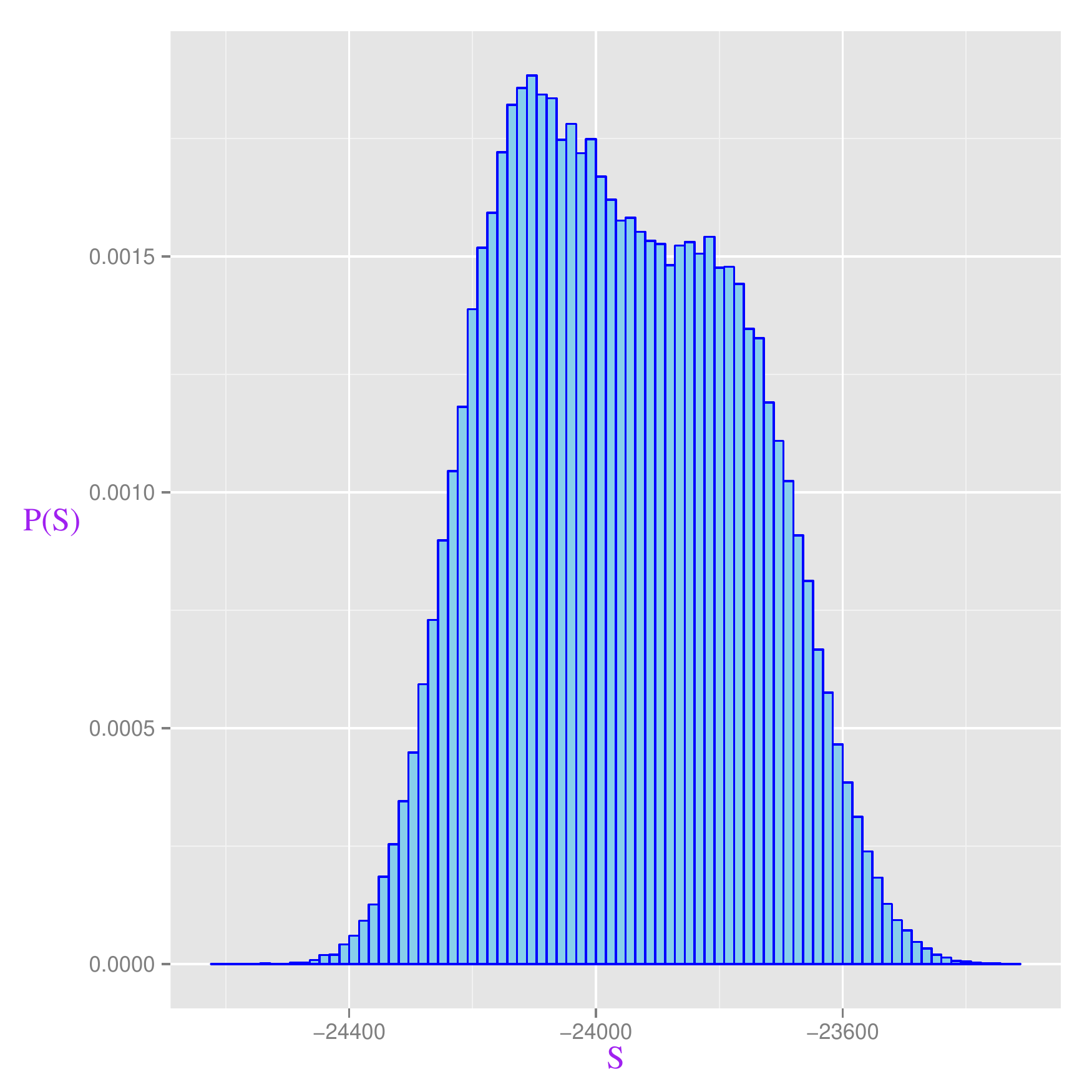} }}}
\caption{}
\label{fig:7}
\end{figure}

\begin{figure}[H]
\centering
\includegraphics[width=3.5in]{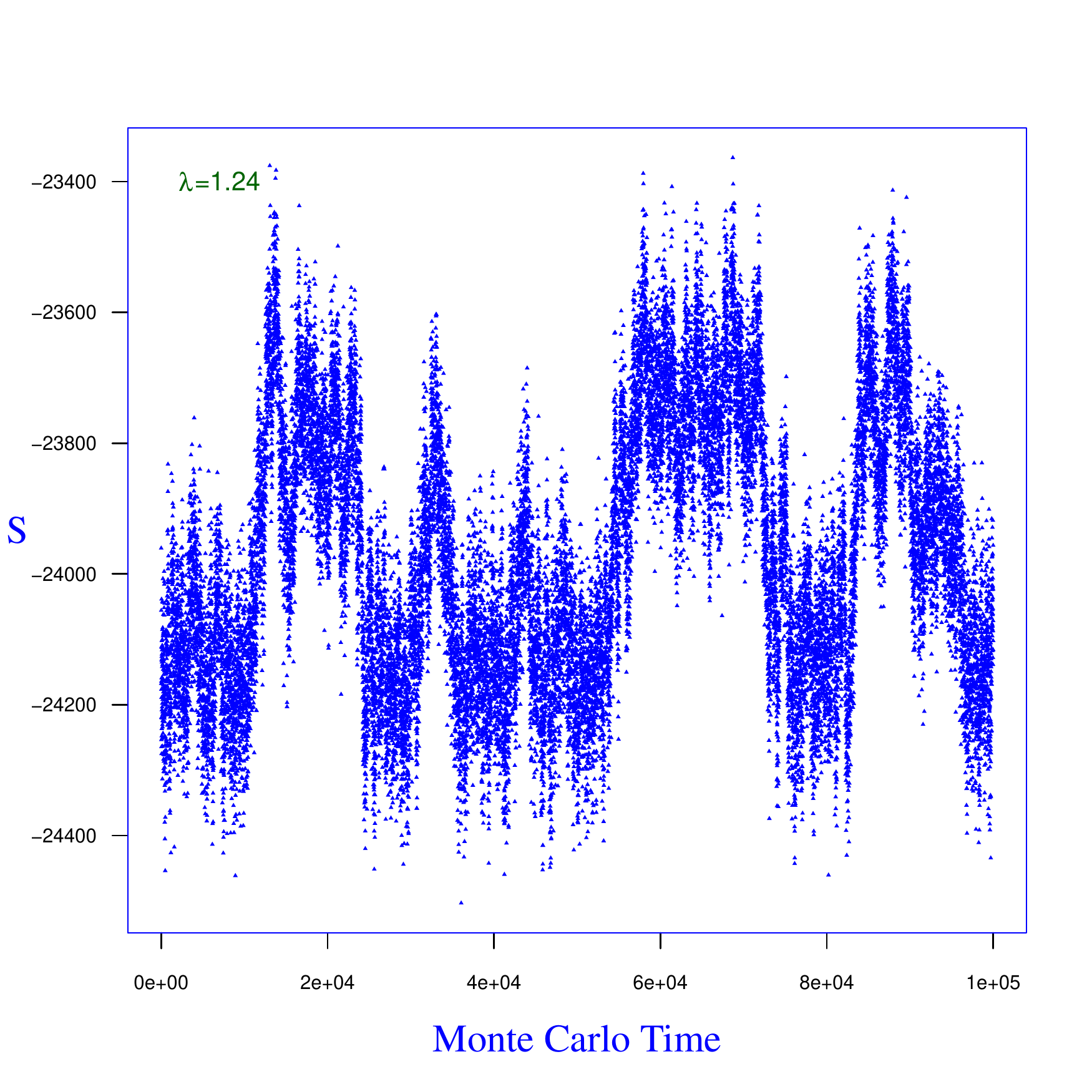}
\caption{}
\label{fig:9}
\end{figure}

In this paper we perform, the first and second analyses. In FIG. (\ref{fig:9}), our Monte Carlo action data ($N=12$ and $L=13$ and $\lambda=1.24$) is plotted against the Monte Carlo time. This value of $\lambda$ is close to its critical value, as confirmed by a simple visual inspection of the plots in FIG.(\ref{fig:1}). The advertised tunneling events can be clearly seen. The plots in FIG.(\ref{fig:7}) display the action histograms for $\lambda=1.24, 1.23$ both close to $\lambda_c$ for system size $L=13$ with $N=12$. They show a bimodal structure. As mentioned earlier, this is the hallmark of a first-order transition.  

\begin{figure}[H]
\centering
\mbox{\subfigure{\includegraphics[width=3.0in]{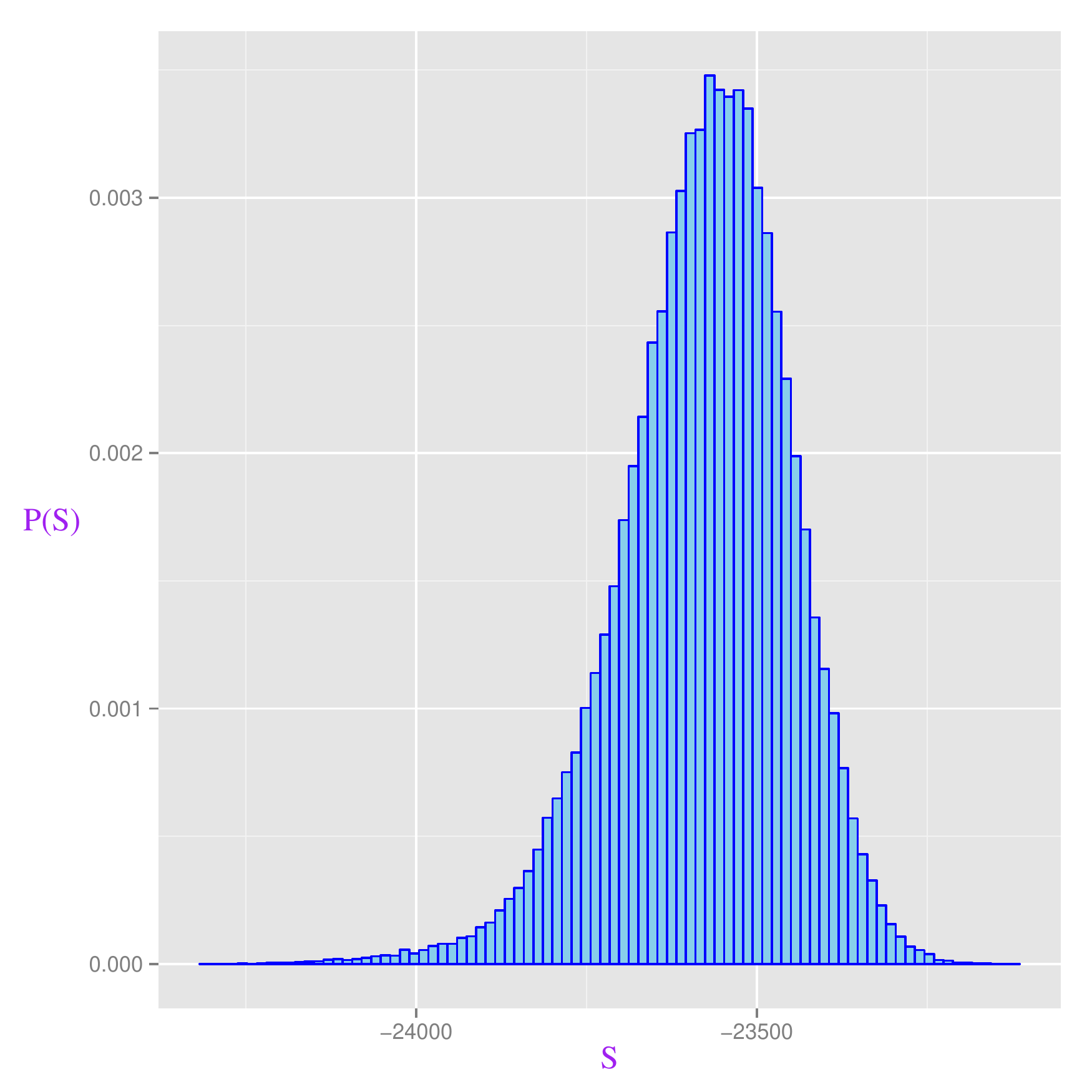}
\quad
\subfigure{\includegraphics[width=3.0in]{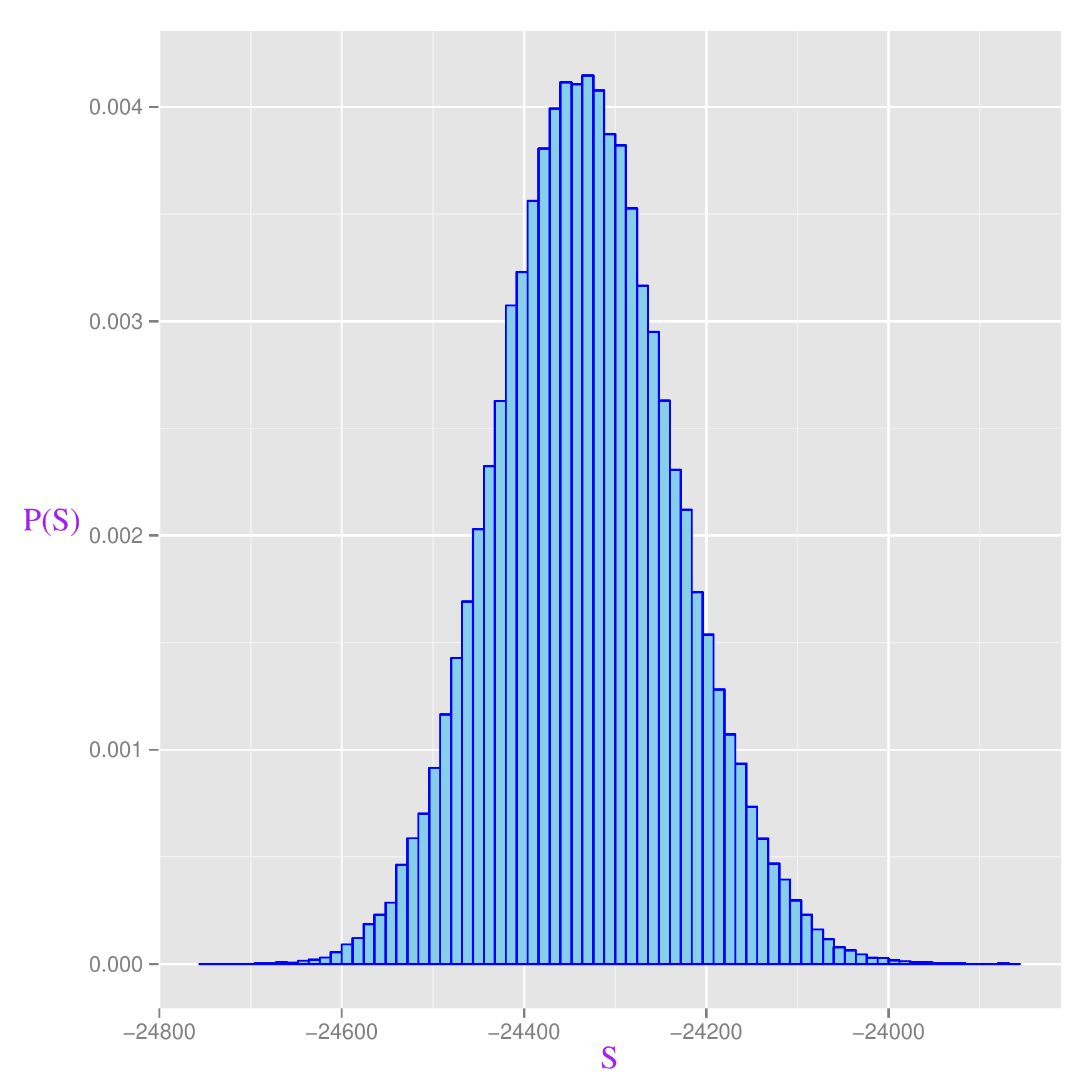} }}}
\caption{}
\label{fig:8}
\end{figure}

Now let us compare the histograms in FIG.(\ref{fig:7}) to histograms for two off-critical $\lambda$ values. In FIG.(\ref{fig:8}), the action histograms for the same values of $N$ and $L$ is presented but for $\lambda=1.21$ (left plot, strongly coupled) and $\lambda=1.28$ (right plot, in the deconfining phase). The bimodal feature disappears and one obtains (slightly skewed) Gaussian normal-looking distributions. Note that the mean of the two normal distributions are not identical. This is to be expected, since the two dominating configurations in the two phases have different actions.
 
%%%%%%%%%%%%%%%%%%%%%
 \section{Summary and outlook}
%%%%%%%%%%%%%%%%%%%%%

In this work phase diagram of the $U(N)$ matrix quantum mechanics was studied using Monte Carlo method at finite-$N$ and finite volume. We confirmed that this system undergoes a phase transition in line with the large-$N$ and infinite-volume predictions of \cite{Gordon1, Gordon2}. To characterize the phase transition, critical behavior of the order parameter fluctuations, i.e., the Polyakov Loop Susceptibility was carefully examined. The transition critical coupling computed at finite-volume was extrapolated to the infinite volume. Based on Monte Carlo time series plot and action histograms, we provided evidence that the transition is first-order. 

In a work in preparation a more thorough analysis of the UMQM thermodynamics will be presented, where the free energy density and latent heat are computed using Monte Carlo. A more complete characterization of the finite-size effects will also be reported. 

This system could also serve as a toy example for modeling disorder in the context of large-$N$ gauge theories, where ab initio studies of its effects can be performed. The fugacity coefficient in this model can be promoted to a non-uniform function of the system spatial extent. Monte Carlo simulations performed here can be run for the new setup without much change. It will be interesting to see if the impact on the thermodynamics is similar to the effects argued for by the author of the present work and collaborators in \cite{Omid1} using holography. We will report on the new results shortly \cite{Omid2}.

\section*{Acknowledgement}

O.S. would like to thank Mithat Unsal for email correspondence and Gordon Semenoff for discussions, the WestGrid high performance computing facility for providing computational resources and Roman Baranowski for assistance. The work of O.S. is supported by the Natural Sciences and Engineering Research Council of Canada (NSERC).

 \appendix

%%%%%%%%%%%%%%%%%%%%%
\section{Metropolis-Hasting algorithm}
%%%%%%%%%%%%%%%%%%%%%
In Metropolis-Hasting local update (realization of a Markov chain), a local update to the field configuration is proposed at a single lattice site during each Monte Carlo time step. This is then repeated throughout the entire lattice. At each step, if the proposed update lowers the action it is accepted with probability one, otherwise it is accepted with probability $e^{-\Delta S}$, where $\Delta S=S[U_{new}]-S[U_{old}]$. To propose updates a set $X=\{x_j\}$ of $5\times10^4$ random unitary matrices in the neighborhood of identity was constructed. This was done by expanding $\exp(i\delta H)$ to sufficiently high order in $\delta$, where $H$ is a random Hermitian matrix. The ``range'' parameter $\delta$ controls  the ``distance'' between this matrix and the identity. At the beginning of each run, using the Newton-Raphson algorithm, we identified the optimal value for the range parameter, such that the acceptance rate for the proposed updates remained around $50\%$. To propose an update $U_i\rightarrow xU_i $ at a given site $i$, a single unitary matrix $x$ was randomly drawn out of the set $X$. The set $X$ was updated from time to time. 
%%%%%%%%%%%
\section{Statistical data analysis}
%%%%%%%%%%%
Statistical data analysis for this project was done using the open source statistical programming language $R$ \cite{R}. Implementation of the algorithms in this paper are in the form of object-oriented C++ and Java. To see the source codes, refer to the author's GitHub repository \cite{GitHub}. Simulations were performed on the WestGrid high performance computing facility.  
%%%%%%%%%%%
\subsection{\small Autocorrelations and Jackknifing}
%%%%%%%%%%%
The unbiased estimator of the expected values we compute, is the arithmetic average of the Monte Carlo measurements along the Monte Carlo time. In fact, every data point has the same mean and variance. Since they are approximately independent samples drawn out of an identical underlying distribution
 \be
\hat{\mathcal{O}}=\frac{1}{p}\sum^{p}_{i=1}\mathcal{O}_{i} =\langle \mathcal{O}\rangle,
\ee
where $p$ is the number of samples (measurements). The unbiased estimator of the variance of the mean is  $\sigma^2_{\hat{\mathcal{O}}}=\sum^{p}_{i=1}(\mathcal{O}_i-\hat{O})^2/p(p-1)$. The presence of autocorrelations in the Monte Carlo time series modifies the variance of $\mathcal{O}$. The actual error estimate when there are autocorrelations is given by 
\be
\langle\mathcal{O}\rangle=\hat{\mathcal{O}}\pm\sigma_{\hat{\mathcal{O}}}\tau_{int},
\ee
where $\tau_{int}$ is the integrated autocorrelation time 
\be\label{autocor}
\tau_{int}=1+2\sum^{\mathcal{N}}_{i=1}\Gamma{(i)}, \quad \frac{C(\mathcal{O}_{i}, \mathcal{O}_{i+\tau})}{C(\mathcal{O}_i, \mathcal{O}_i)}=\Gamma(\tau),
\ee
where $C$ stands for the normalized autocorrelation of the time series. The long-time behavior of $\Gamma=\Gamma(t)$ is erratic. Therefore, time separation in the sum in (\ref{autocor}) needs to be cutoff at some $\mathcal{N}$. Following the common guidelines, we take $\mathcal{N}=5\tau_{exp}$, where $\Gamma(t)\approx\exp(-t/\tau_{exp})$ for short time separations.\footnote{Computing autocorrelations for long Markov chains could be computationally expensive. The short-cut is to assume the exponential falloff holds at all times. In this case one gets $\tau_{int}\approx 1+ 2\tau_{exp}$. Note that to compute $\tau_{exp}$, one only needs to know $\Gamma=\Gamma(t)$ for short time separations.}  

In this paper, for every data point in the plots, we fit an exponential form to the short-time behavior for $C$. After reading off $\tau_{exp}$, we perform the sum in (\ref{autocor}) up to $5\tau_{exp}$. The integrated autocorrelations were then used to form statistically independent bins. The error-bars for each point were computed from the statistically independent bins, using the Jackknife resampling technique \cite{LatticeBook}.  
%%%%%%%%%%%%%
\subsection{\small Confidence boundaries}
The boundaries of the confidence region in FIG.(\ref{fig:3}), FIG.(\ref{fig:4}) and FIG.(\ref{fig:5}) were computed as follows. The prediction error $\sigma_{pred}$ for the quadratic fit at a given point $\lambda$ is given by
\be\label{error}
\sigma^2_{pred}(\lambda)=\sigma^2_{\alpha}\lambda^4+\sigma^2_{\beta}\lambda^2+\sigma^2_{\eta}+2\Cov(\alpha, \beta)\lambda^3+2\Cov(\alpha, \eta)\lambda^2+2\Cov(\beta, \eta)\lambda, 
\ee
where variance of the regression coefficients as well the covariance matrix in (\ref{error}) were estimated by sampling from the (normal) joint distribution function, 
modeled by our data points and their error-bars. The confidence boundaries were then reported as $C_{u, \ell}(\lambda)=F(\lambda)\pm\sigma_{pred}(\lambda)$, where $C_{u, \ell}$ denotes the upper and lower confidence boundaries and $F=F(\lambda)$ is the quadratic fit.  

\end{document}